\documentstyle[pre,aps,epsf]{revtex}

\newcommand{\be}{\begin{equation}}
\newcommand{\ee}{\end{equation}}

\newcommand{\x}{{\bf x}}
\newcommand{\bftheta}{{\mbox{\boldmath $\theta$}}}

\begin{document}

\title{A lattice-Boltzmann model for interacting amphiphilic fluids}
\author{Maziar Nekovee and Peter V. Coveney}
\address{Centre for Computational Science, \\
Queen Mary and Westfield College, University of London,\\
Mile End Road, London E1 4NS, U.K.}
\author{Hudong Chen}
\address{Exa Corporation,\\ Lexington, Massachusetts, U.S.A.}
\author{Bruce M. Boghosian}
\address{Center for Computational Science, \\
Boston University, 3 Cummington Street, Boston, Massachusetts
02215, U.S.A.}

\maketitle

\begin{abstract}
{We  develop our recently proposed lattice-Boltzmann method for 
the non-equilibrium dynamics of amphiphilic fluids \cite{cbc}.
Our method maintains an orientational degree of freedom for the
amphiphilic species and models fluid interactions at a microscopic
level by introducing self-consistent mean-field forces between 
the particles into the lattice-Boltzmann dynamics, 
in a way that is consistent with kinetic theory. We present 
the results of extensive simulations in two dimensions 
which demonstrate the ability of our model to capture the 
correct phenomenology of binary and ternary amphiphilic fluids.}

\end{abstract}



\section{Introduction}
The lattice-Boltzmann (LB) method is a mesoscopic approach to 
the study of the dynamics of fluids, intermediate between macroscopic 
models, based on Navier-Stokes equations, and 
microscopic approaches, based on molecular dynamics (MD) simulations.
In this method, time, velocity and space are all discretized 
and a single-particle distribution function is utilized to describe
the time evolution of an ensemble of molecules having a discrete
set of possible velocities \cite{review,succi}.
The computationally demanding tracking of individual molecules is thus
avoided and at the same time 
the ensemble-averaged distribution function retains much of the 
microscopic information. Macroscopic or hydrodynamic effects 
naturally emerge from mesoscale lattice-Boltzmann dynamics, 
provided that the LB  equation possesses the correct and necessary 
conservation laws and symmetries \cite{review}. 
Historically, the lattice-Boltzmann model  
was first derived  from its predecessor--the lattice-gas
automata (LGA) \cite{lga1,lga2,lga3,rothman}-- 
but it has been shown recently that the model 
can also be derived from kinetic theory by discretization of the 
Boltzmann equation in velocity, space and time \cite{he1,he2}.

An important and promising area of application of the lattice-Boltzmann
method is for modeling the dynamics of multicomponent
fluids. There is in particular much fundamental and
technological interest in modeling amphiphilic systems; 
i.e. fluids comprising 
one or two immiscible phases (such as oil and water) and an amphiphilic 
species (surfactant). The addition of surfactant to a binary immiscible
fluid can result in complex structures on a mesoscopic length scale,
including lamellar, micelles and microemulsion phases \cite{gelbart,gompsh}.
The non-equilibrium dynamics and hydrodynamics 
of these systems are difficult to simulate using a
continuum approach based on Navier-Stokes equations. 
One major difficulty these methods face is the existence of complicated 
interfaces between fluid components which can undergo topological 
changes; another is that it is far from clear how to formulate a
hydrodynamic description of these fluids.
Microscopic  approaches, based on molecular dynamics,  are able 
to deal with amphiphilic fluids, but they are still computationally
too  demanding to investigate large-time
dynamics (times scales accessible to MD are typically of the order 
of nanoseconds \cite{jb} whereas important 
time scales in such complex fluids 
vary between $10^{-11}-10^{4}$ seconds) 
and the extended spatial structures involved in many problems of interest.

The formation of fluid interfaces is a consequence 
of interaction between the molecules of the fluids
\cite{rowlinson}, and the LB models for
multicomponent fluids come in two varieties, depending on the way 
interactions between fluid components are incorporated.
In the free-energy approach [13-16]
the starting point is an Ansatz for the free-energy functional 
of the complex interacting fluid and the approach to equilibrium
is assumed to be governed by a Ginzburg-Landau or Cahn-Hillard
equation or some suitable generalization thereof.
A LB collision operator is then constructed that gives 
rise to the desired evolution equation in the hydrodynamic 
limit. The most attractive feature of this model is that the 
equilibrium state is fixed {\it a priori} by the choice of free-energy
functional. This guarantees thermodynamic consistency 
at equilibrium but limits the description 
of non-equilibrium dynamics. For example, the one-component 
free energy model is not Galilean invariant \cite{julia2}. 
Furthermore, the method relies on the ability to postulate a suitable form 
for the free energy, which might not be always available.
An alternative approach, which is closer in spirit to kinetic 
theory, is to add interactions between fluid components 
by introducing intermolecular forces 
between fluid particles \cite{ch,he3}.
Very recently we proposed such a  lattice-Boltzmann model for 
amphiphilic fluids \cite{cbc} which is based on introducing the dynamics and
interactions of amphiphilic molecules 
into the multicomponent lattice-Boltzmann scheme of Shan and Chen 
\cite{ch}. In the resulting ternary vector model, which was 
inspired by the lattice-gas automata (LGA) formalism of Boghosian 
Coveney and Emerton \cite{emerton1},
amphiphilic species possess both translational
and orientational degrees of freedom, a feature which is crucial
for capturing much of the above-described phenomenology of 
amphiphilic systems.

The principal aim of the present paper is to further develop and
explore this model and to assess and establish its general validity for
modeling the dynamics of amphiphilic systems. In Section II we give 
a detailed description of the model and show that 
with slight modification, the  mean-field 
treatment of intermolecular interactions in the model \cite{cbc,ch} 
can be made 
consistent with the mean-field kinetic theory of interacting fluids, 
based on a set of coupled Boltzmann equations. 
In section III we use this modified model to simulate 
the non-equilibrium dynamics of binary immiscible and binary and 
ternary amphiphilic fluids in two-dimensions. 
These simulations demonstrate that 
the model exhibits the correct dynamical behavior in the case
of critical quenches of binary fluids and it is capable 
of describing in a consistent way the phenomenology of various
experimentally observed self-assembling 
structures,  such as lamellae, droplet and sponge microemulsion phases,
the arrest of separation of immiscible oil-water phases
when enough surfactant is present in the system, and the formation 
of the lamellar phases in binary water-surfactant systems.
Finally, in section V we draw some conclusions from 
this work and discuss the prospects of future developments of the model.
\section{Lattice-Boltzmann model for interacting fluid mixtures}
The lattice-Boltzmann scheme for a fluid mixture having 
$S$ components is defined 
in terms of the distribution functions $f_i^\sigma(\x,t)$ belonging to
the particles of component $\sigma$ of the fluid mixture 
(e.g. $\sigma =$ water or oil in a binary system).

In the BGK approximation \cite{bgk} of the collision operator 
these equations are \cite{succi,he1}
\be
f^\sigma_i({\bf x} + {\bf  c}_i\Delta t, t + \Delta t)- 
f^\sigma_i({\bf x}, t) = -
\frac{f^\sigma_i-f_i^{\sigma (eq)} }{\lambda_\sigma} \Delta t
\label{eq:mboz}
\ee
where ${\bf c}_i$ $(i=0,1 \ldots M)$ are a set of discrete velocity 
vectors on a regular lattice, 
${\bf x}$ is a point on the underlying spatial grid, 
$f_i^{\sigma (eq)}$ is a local equilibrium distribution
function, $\lambda_{\sigma}$ is the relaxation time for component 
$\sigma$ and  $\Delta t$ is the timestep.

The number density 
$n^\sigma$ and the mean velocity ${\bf u}^\sigma$ of component 
$\sigma$ are given by 
\be
n^{\sigma}(\x,t) = \sum_i \; f_i^{\sigma},
\ee
\be 
\rho^{\sigma}(\x,t) u^{\sigma}(\x,t) = m^\sigma  
\sum_i \; {\bf c}_i f_i^{\sigma}, 
\ee
with $\rho^{\sigma} = m^{\sigma}n^{\sigma}$ the density 
and $m^{\sigma}$  the  mass of species $\sigma$.

The equilibrium distribution function should be chosen 
in such a way that the BGK collision operator $\Omega^\sigma_i =
-(f_i^\sigma-f_i^{\sigma (eq)})/\lambda_\sigma$ locally conserves the 
mass of each species $\sigma$
\be 
\sum_i \Omega_i^ \sigma = 0. 
\ee
We  also require that the momentum of all components together should
be conserved locally 
\be
\sum_\sigma m_\sigma \sum_i {\bf c}_i \Omega^\sigma_i = 0.
\ee
A common choice which satisfies the above constraints is 
\cite{chen4,qian}
\be
f_i^{\sigma (eq)} = \omega_i n^{\sigma}\left [ 1 + \frac{1}{c_s^2} 
{\bf c}_i \cdot {\tilde{\bf u}} + 
\frac{1}{2c_s^4} ({\bf c}_i \cdot {\tilde{\bf u}})^2 - \frac{1}{2c_s^2} 
\tilde{u}^2 \right],
\label{eq:equildist}
\ee
where $\omega_i$ is a weight factor and $c_s$ is the speed of sound,
both of which are determined entirely by the choice of lattice.
For example, in the so-called D3Q19 lattice \cite{qian} (19 velocity
vectors in three dimensions) $c_s=1/\sqrt{3}$ and 
$\omega_i=1/3$, $1/18$ and
$1/36$ for $e_i = 0$, $e_i= 1$ and $e_i =2$, respectively.

The requirement that the total velocity of all components 
together should be conserved at each lattice point results 
in the following form for the common velocity  ${\tilde {\bf u}}$
\be 
 {\tilde {\bf u}} = \frac {\sum_\sigma \frac {\rho^\sigma {\bf u}^\sigma}
{\tau_\sigma} } { \sum_\sigma \frac {\rho^\sigma} {\tau_\sigma}}. 
\ee
In the limit of small Mach numbers, i.e. $\tilde{u}/c_s\ll1$, 
where (6) 
is positive, the above choice of $f_i^{\sigma (eq)}$ results 
in Navier-Stokes hydrodynamics, where each fluid component $\sigma$ 
has kinematic viscosity 
$\nu_{\sigma} = (\lambda_\sigma/\Delta t - 1/2)c_s^2$
\cite{chen4}.

\subsection*{Mean-field treatment of interactions} 

In order to describe immiscibility effects, some form of repulsive 
interaction between fluid components must be introduced into the LB
equations. In the Shan-Chen scheme \cite{ch}, coupling between 
fluid components is achieved by including intermolecular
interactions in the model. These interactions are modeled as
a self-consistently generated mean-field force.
Allowing only homogeneous isotropic interactions between
nearest-neighbors, the mean-field force ${\bf F}^{\sigma}$  acting on
particles of component $\sigma$  is given by
\be
{\bf F}^\sigma ({\bf x},t) = - \psi^\sigma ({\bf x},t)
\sum_{\bar \sigma} \sum_{{\bf x}'} g_{\sigma 
{\bar \sigma}} \psi^{\bar \sigma} ({\bf x}',t) ({\bf x}' - {\bf x})
\label{eq:force}
\ee
where $\psi^\sigma= \psi^\sigma(\rho^{\sigma}({\bf x},t))$ 
is the so-called effective mass, which can have a general form for
modeling various types of fluids (in the present study we have taken 
$\psi^\sigma=\rho^{\sigma}$) and $g_{\sigma {\bar \sigma}}$ is 
a force coupling constant, whose magnitude controls the strength of 
the interaction between component $\sigma$ and ${\bar{\sigma}}$ and 
whose sign determines whether this interaction is repulsive 
($g_{\sigma {\bar \sigma}}>0$) or attractive ($g_{\sigma {\bar \sigma}}<0$).
Shan and Chen incorporated the above force term in the LB dynamics 
by adding an increment \cite{ch}
\be
\delta {\bf u}^{\sigma} = 
\frac{{\bf F}^{\sigma}}{\rho^\sigma}  \tau_\sigma \Delta t
\ee
to the velocity $\tilde{{\bf u}}$ which enters the equilibrium 
distribution function (6) in the BGK collision operator.

With the choice $\psi^\sigma=\rho^{\sigma}$ the above treatment of 
intermolecular interactions is consistent with kinetic theories of
interacting fluids, in which  non-local interactions are described by 
including a self-consistently generated mean-field force term in the 
Boltzmann equation while local interactions are treated as true
collisions. Indeed, kinetic equations of this type 
have been successfully applied to describe
gas-gas separation into two phases \cite{boltz1} and, more recently,
spinodal decomposition in binary fluids \cite{boltz2}. 
However, the way the mean-field force enters the Shan-Chen model 
is not entirely consistent with the Boltzmann equation in the presence
of a force \cite{luo,martys}.
The explicit expression for the force term in the Shan-Chen model
can be obtained by substituting ${\tilde {\bf u}}+\delta {\bf u}$ into the 
equilibrium distribution function. This yields 

\begin{eqnarray}
f^\sigma_i({\bf x} + {\bf  c}_i\Delta t, t + \Delta t) - 
f^\sigma_i({\bf x}, t) = - \frac{f^\sigma_i-f_i^{\sigma_i (eq)} }
{\tau_\sigma}
&+&  \omega_i n^\sigma \left[
\frac{ {\bf c}_i - {\tilde {\bf u}} }{c_s^2} + 
\frac{ ({\bf c}_i. {\tilde {\bf u}} )}{c_s^4}{\bf c}_i \right] .
{\bf a}_\sigma \Delta t \nonumber \\
&-&  \frac{1}{2} \omega_i n^\sigma \left[ \frac{a_\sigma ^2}{c_s^2}- 
\frac{({\bf c}_i.{\bf a_\sigma})^2}{c_s^4} \right] \tau_\sigma 
\Delta t^2.
\end{eqnarray}
where ${\bf a}_{\sigma} = {\bf F}^\sigma/\rho^\sigma $ is force per 
unit mass and we have introduced the dimensionless relaxation time 
$\tau_\sigma = \lambda_\sigma/\Delta t$.
As was shown by Luo \cite{luo} and by Martys {\it et al.}
\cite{martys}, 
by omitting the second order term in ${\bf a}_\sigma$ 
the above equation can be made consistent
with a systematic derivation of the lattice-Boltzmann model in the 
presence of the  force ${\bf a}_{\sigma}$, as obtained from
discretization of the corresponding Boltzmann equation, 
in the low Mach-number regime.
In our scheme we omit this non-linear 
term when calculating the effect of the self-consistently generated 
mean-field force, so that the resulting lattice-Boltzmann scheme is 
consistent with the underlying Boltzmann equation in the presence
of a force. Our final lattice-Boltzmann equation is then given by 
\begin{eqnarray}
f^\sigma_i({\bf x} + {\bf  c}_i\Delta t, t + \Delta t) - 
f^\sigma_i({\bf x}, t) = - \frac{f^\sigma_i-f_i^{\sigma_i (eq)} }
{\tau_\sigma}
&+&  \omega_i n^\sigma \left[
\frac{ {\bf c}_i - {\tilde {\bf u}} }{c_s^2} + 
\frac{ ({\bf c}_i. {\tilde {\bf u}} )}{c_s^4}{\bf c}_i \right] \cdot
{\bf a}_\sigma \Delta t
\label{eq:mazboltz} 
\end{eqnarray}

In the case of a single-component interacting fluid rewriting 
the force term in terms of distribution functions 
$n_i^{\sigma}$ leads to  the following lattice-Boltzmann equation
\be
f^\sigma_i({\bf x} + {\bf  c}_i\Delta t, t + \Delta t) - 
f^\sigma_i({\bf x}, t) = 
-\frac {f^\sigma_i-f_i^{\sigma_i (eq)} }{\tau_\sigma}
+ \sum_j \Lambda_{ij}^{\sigma \sigma}f_j^{\sigma}
\ee
where $\Lambda_{ij}^{\sigma \sigma}$ is given by 
\be
\Lambda_{ij}^{\sigma \sigma} = \omega_i \left[\frac{1}{c_s^2}
 ({\bf c}_i-{\bf c}_j)+
\frac{{\bf c}_i \cdot {\bf c}_j}{c_s^4} {\bf c}_i\right].
{\bf a}_\sigma \Delta t 
\ee
Thus the net effect of the force term  is to 
introduce off-diagonal matrix elements in the BGK collision operators. 
These matrix elements describe transitions 
from state ${\bf c}_i$ to state ${\bf c}_j$, and {\it vice versa}, due 
to the drift of particles under the influence of 
the force during the time $\Delta t$. In the case of a fluid mixture
the BGK collision operator couples different fluid components
even in the absence of the mean-filed force, through the 
common velocity $\tilde{\bf u}$ which enters $f_i^{\sigma (eq)}$.
The mean-field force introduces an additional non-local 
coupling between fluid particles and in this case 
Eq. (11) can be rewritten as
\be
f^\sigma_i({\bf x} + {\bf  c}_i\Delta t, t + \Delta t) - 
f^\sigma_i({\bf x}, t) = 
-\frac {f^\sigma_i-f_i^{\sigma_i (eq)} }{\tau_\sigma}
+ \sum_{{\bar{\sigma}}}\sum_j \Lambda_{ij}^{\sigma {\bar{\sigma}}}
f_j^{{\bar{\sigma}}}
\ee
where $\Lambda_{ij}^{\sigma {\bar{\sigma}}}$ are matrix elements of the 
cross-collision operator 
\be
\Lambda_{ij}^{\sigma {\bar{\sigma}}} = \omega_i 
\left[\frac{1}{c_s^2}
 (\delta_{\sigma {\bar{\sigma}}}{\bf c}_i- \alpha_{\sigma {\bar{\sigma}}}
{\bf c}_j)+ \alpha_{\sigma,{\bar{\sigma}}}
\frac{{\bf c}_i.{\bf c_j}}{c_s^4} {\bf c}_i\right].
{\bf a}_\sigma \Delta t 
\ee
with 
\be
\alpha_{\sigma {\bar{\sigma}}} = \frac{n^\sigma}{n^{{\bar{\sigma}}}} \times
\frac {\frac {\rho^{{\bar{\sigma}}}}
{\tau_{{\bar{\sigma}}}} } { \sum_{{\bar{\sigma}}} \frac {\rho^{{\bar{\sigma}}}} {\tau_{{\bar{\sigma}}}}}
\ee

The above treatment of intermolecular interactions does not take 
into account strong repulsive interactions between molecules 
at short distance which prevent them to overlay 
(the excluded volume effect \cite{enskog}).
This effect is important in calculating transport coefficients of 
dense gases and fluids \cite{enskog} but we do not expect it to significantly 
alter the dynamics of microemulsion, which is the main focus of our 
present work. The excluded volume  effect can be introduced in the 
model by adding Enskog corrections to the BGK collision operator \cite{he3}.

\subsection{Modeling surfactant dynamics and interactions}
A surfactant (or amphiphilic) molecule usually possesses two different 
fragments, each having an affinity for 
one of the two immiscible components. In the case of a prototype 
mixture of immiscible fluids like oil and water, a typical surfactant 
molecule would be an amphiphile that has a hydrophilic head
preferring to be in contact with water molecules, and a hydrophobic 
tail preferring to be surrounded by oil. Under equilibrium conditions 
the surfactants are predominantly adsorbed at the oil-water 
interface, hence effectively screening the two-body repulsive
interaction between 
immiscible species.
In our scheme \cite{cbc} these
essential characteristics of amphiphilic molecules are modeled by introducing
a separate amphiphilic species into the Shan-Chen model which possesses
both translational and rotational degrees of freedom. This is achieved
by modeling these species as idealized point dipoles.
Consequently, the interaction of 
the amphiphilic molecules with each other and with non-amphiphilic 
molecules (like oil and water)
depends not only on their relative distance but also on their dipolar 
orientation \cite{emerton1}.
Furthermore, a full description of the dynamics of amphiphilic
molecules requires a description of the propagation of each single-particle 
distribution function as well as the time evolution of 
their dipole vectors. Making the physically
reasonable assumption that the dipole moment carried by amphiphilic 
molecules is independent of their velocity, an average
dipole vector ${\bf d}(\x,t)$ is introduced at each site
$\x$ representing the orientation of any amphiphile present there. 
The propagation of the amphiphilic molecules  is then described 
in our scheme by 
a set of BGK-like equations, one for the distribution 
function $f_i^{s}$ and one for the relaxation of the average dipole 
vector ${\bf d}(\x,t)$ to its local equilibrium 
orientation \cite{cbc}:
\be
f^s_i({\bf x} + {\bf  c}_i\Delta t, t + \Delta t)-
f^s_i({\bf x}, t) =
-\frac{f^s_i-f^{s (eq)}_i}{\tau_s}  
+ \sum_{\sigma}\sum_j \Lambda_{ij}^{\sigma s}f_j^\sigma + 
\sum_j \Lambda_{ij}^{ss}f_j^s,
\ee
\be
{\bf d}(\x,t+\Delta t) = \bar{{\bf d}}(\x,t) - \frac{1}{\tau_d}
[\bar{{\bf d}}(\x,t)-{\bf d}^{eq}(\x,t)].
\ee
In the above equations $f^s_i(\x,t)$ is the distribution function of 
surfactant molecules and  $\tau_s$ and $\tau_d$ are
dimensionless relaxation times representing the 
rates of relaxation of $f^s_i(\x,t)$ and ${\bf d}(\x,t)$ to their local 
equilibrium distributions  $f^{s (eq)}_i$ and ${\bf d}^{eq}(\x,t)$,
respectively. 
The collision operators $\Lambda_{ij}^{\sigma s}$ and 
$\Lambda_{ij}^{ss}$ have the same form as (15) and describe 
the effect of the total mean-field force experienced by surfactant 
particles due to their interactions with non-amphiphilic particles and
with other amphiphilic particles. We shall specify 
these forces shortly.
Finally, $\bar{{\bf d}}(\x,t)$ is obtained from 
the following equation of motion
\be
n^s(\x,t){\bar {\bf d}}(\x,t) = \sum_{i} f_i^s(\x-
{\bf c}_i \Delta t){\bf d}(\x-{\bf c}_i\Delta t ,t).
\ee

The equilibrium distribution function $f^{s (eq)}_i$ is chosen to have
the same form as ~(\ref{eq:equildist}) and, 
in analogy with the Weiss molecular field 
theory of magnetism \cite{weiss}, the 
equilibrium distribution ${\bf d}^{eq}(\x,t)$ is obtained self-consistently
from the Boltzmann distribution as
\be
{\bf d}^{eq}(\x,t) = d_0 
\frac { \int e^{-\beta U} \hat{{\bf v}} d \Omega }
{\int e^{-\beta U} d \Omega },
\label{eq:dipole}
\ee
where
$d \Omega$ is an element of solid angle, 
${\hat {\bf v}}$ is a unit vector and 
$U$ is the potential energy of a dipole in the 
mean field generated by dipolar amphiphiles and non-amphiphilic molecules:
\be
U = - \hat{{\bf v}}\cdot({\bf h}_s +{\bf h}_c).
\label{eq:energy}
\ee
In ~(\ref{eq:energy}) ${\bf h}_s$ and ${\bf h}_c$ are the mean fields
resulting from dipole-dipole
interactions and the interaction between surfactant dipoles and 
non-amphiphilic molecules, respectively. In Eq. ~(\ref{eq:dipole})
both $d_0$ and the inverse 
temperature $\beta$, are independent parameters.
Performing the integral yields for the equilibrium distribution 
\be
{\bf d}^{eq} = d_0  
\left[\coth\left(\beta h \right)
- \frac {1} {\beta h} \right] {\hat {\bf h}} 
\ee
in 3D, and 
\be
{\bf d}^{eq} = d_0 \left[
\frac{I_1(\beta h)}{I_0(\beta h)} \right] \hat{{\bf h}} 
\ee
in 2D, where $\hat{{\bf h}}= {\bf h}/h$ and
$I_0$ and $I_1$ are the zero and the first order modified Bessel
functions \cite{arfken}, respectively. 

The mean-field generated by water and oil molecules is given by
\be
{\bf h}^c ({\bf x}, t) =
\sum_\sigma e_\sigma \sum_i n^\sigma ({\bf x} + {\bf c}_i \Delta t, t){\bf c}_i
\ee
where $e_\sigma$ is the ``charge'' for various molecular components
(which may take its values from the set $\{-1,0,1\}$). In the 
present simulations we take $e=1$ for water molecules and $e=-1$ for
oil molecules. Similarly, the mean-field generated by other surfactant
molecules is given by
\be
{\bf h}^s ({\bf x}, t) 
= \sum_i [\sum_{j \neq 0} 
n^s_i({\bf x} + {\bf c}_j \Delta t , t) \bftheta_j \cdot
{\bf d}_i({\bf x} + {\bf c}_j \Delta t,t)
+  n^s_i ({\bf x}, t){\bf d}_i({\bf x}, t)],
\label{hs}
\ee
where 
\[ \bftheta_j = {\bf I} - D \frac {{\bf c}_j {\bf c}_j} {c_j^2} \]
is the traceless second-rank tensor, and $D$ is dimension of the 
lattice.

Finally, we specify the form of interaction between amphiphilic 
molecules with water and oil particles and between amphiphilic 
molecules themselves. These are obtained from Eq. ~(\ref{eq:force})
by treating
each amphiphilic molecule as a pair of water and oil molecules displaced
by a distance ${\bf d}(\x,t)$ from each other and performing a Taylor 
expansion in ${\bf d}$ in the resulting expression for the total force
\cite{cbc}.
Assuming that the dipole head and tail have equal and opposite 
charges $e=\pm1$ and only nearest-neighbor interactions 
considered, the additional forces are given by

\be
{\bf F}^{\sigma , s}({\bf x},t)
= - 2\psi^\sigma ({\bf x},t) g_{\sigma s}
\sum_{i\neq 0}{\bf d}({\bf x} + {\bf c}_i  \Delta t)
\cdot ({\bf I} - \frac {{\bf c}_i {\bf c}_i} {c_i^2} D)
\psi^s ({\bf x} + {\bf c}_i  \Delta t,t)
\label{bscn} 
\ee
\be
{\bf F}^{s, c}({\bf x},t)
= 2\psi^s ({\bf x},t) {\bf d}({\bf x},t)
\cdot \sum_\sigma g_{\sigma s}
\sum_{i \neq 0} ({\bf I} - \frac {{\bf c}_i {\bf c}_i} {c_i^2} D)
\psi^\sigma ({\bf x} + {\bf c}_i,t)
\label{bcsn} 
\ee
and 
\begin{eqnarray}
{\bf F}^{s,s}({\bf x},t) &=& - \frac {4D} {c^2}
g_{ss} \psi^s({\bf x}) \sum_i \{
{\bf d}({\bf x} + {\bf c}_i \Delta t,t) {\bf d}({\bf x},t) :
[{\bf I} - \frac {{\bf c}_i {\bf c}_i} {c_i^2} D] {\bf c}_i
\nonumber \\
&+& [ {\bf d}({\bf x} + {\bf c}_i \Delta t,t) {\bf d}({\bf x},t)
+ {\bf d}({\bf x},t) {\bf d}({\bf x} + {\bf c}_i  \Delta t,t)] 
\cdot {\bf c}_i \} \psi^s({\bf x} + {\bf c}_i + \Delta t,t)
\label{eq:nssf}
\end{eqnarray}
In the above equations  
${\bf F}^{\sigma,s}$ is the force acting on 
non-amphiphilic particles $\sigma$ (water and oil) 
due to amphiphile dipoles, ${\bf F}^{s,c}$ is the force 
acting on amphiphilic molecules due to all non-amphiphilic particles and 
${\bf F}^{s,s}$ is the force among amphiphilic molecules themselves. 
The  coupling constants $g_{\sigma s}$ and $g_{s s}$ determine,
respectively, the strength of interaction between water/oil molecules
and surfactant molecules, and among surfactant molecules themselves.
The coupling coefficient $g_{ss}$ should be chosen negative if we wish
to model attraction between two amphiphile heads or tails,
and repulsion between a head and a tail.

\section{simulations}
As mentioned earlier the principal aim of the present work is 
to assess the ability of our model to reproduce the basic properties 
of self-assembling amphiphilic fluids. For this reason 
we choose not to explore the entire parameter space of the model in 
the present study but rather find through a limited search in this 
space a canonical set of 
parameters which allows us to describe generic behavior in a
consistent way. 
Another important consideration in choosing the parameters relates to 
the numerical instabilities which, for a given concentration of 
oil, water and surfactant, were found to occur 
when the force coupling constant $g_{\sigma {\bar{\sigma}}}$, $g_{\sigma s}$
and $g_{ss}$, or 
the mean densities, were increased beyond certain threshold values. 
We found that these instabilities occur when the forcing terms cause
the right-hand side of 
Eqs. (14) and (17) to become negative and are  caused by a combination
of the mean-field treatment of interparticle interactions and the restriction 
of the lattice-Boltzmann scheme to low Mach numbers.
With these considerations in mind, and after a restricted search in
the parameter space of the model 
we arrived at the following  set of canonical parameters 
which, unless stated otherwise, 
are used throughout our simulations (the timestep  $\Delta t$ is set
to $1$ throughout) 
$g_{\sigma \sigma}=0$, $g_{\sigma {\bar{\sigma}}}=0.03$, 
$g_{\sigma s} =-0.01$, $g_{ss} = 0.01$, $\tau_{\sigma}=\tau_s=1$, 
$\tau_d=2$, $m^{\sigma}=1$, $m^{s}=2$, $d_0=1$  and $\beta=10$. All 
calculations were performed on a $256^{2}$ lattice subject to 
periodic boundary conditions. In the case of the binary oil-water systems 
we also performed additional calculations on  $512^{2}$ and
$128^{2}$ lattices in order to check for finite size effects.

The CPU time and memory requirements of our algorithm both scale 
$\sim L^{D}$, where $L$ is the linear dimension of the lattice.
In two dimensions we were able to study adequate system sizes using 
typical workstations. 
For example, simulation of a ternary system on a
$256^2$ lattice on a Silicon Graphics 250 MHz processor
workstation takes just under $8$  hours to reach $3000$ timesteps
(switching off the subroutines which perform the surfactant dynamics 
calculations reduces the CPU time by $50\%$.)
For the present calculations we used a serial implementation of our 
lattice-Boltzmann algorithm. 
In three dimensions, however, the serial algorithm quickly becomes 
prohibitive in terms of computer memory for moderately sized
systems. Fortunately, an important feature of LB is its intrinsically 
parallel structure and we  have implemented a parallel version of 
our algorithm \cite{jon} 
which allows us to perform very large-scale $3D$ simulations
on massively parallel platforms. We plan to report on the three
dimensional model in future publications.
\subsection{Binary Oil-Water System}
The dynamics of phase separation in a binary mixture, following a
thermal quench into the unstable coexistence regime, proceeds by
spinodal decomposition. Immediately after the quench, small domains, 
with local concentrations roughly corresponding to that of the 
two  pure immiscible  phases, spontaneously form and grow and finally 
result in complete phase separation \cite{gompsh}.
To simulate phase separation we set up a simulation with equal quantities
of water and oil with average densities $0.5$, and no surfactant
present (a ``critical quench'').
This choice of average densities 
ensures that we are well within the immiscibility region of the model.
The initial condition of these simulations is a uniform mixture of the 
two fluids with small random fluctuations in the uniform densities.
These fluctuations are necessary to move the system out of a 
metastable uniform state, in which the mean-field forces are identically
zero. The force term  is initially set to zero and
$f_i^{\sigma}$ are set to $f_i^{\sigma (eq)}$ calculated from 
$n^{\sigma}$ and ${\tilde {\bf u}}=0$, (Eq. (7)). 

As can be seen from Fig. 1, immediately after the 
quench small domains spontaneously start to form.
As time evolves, sharp interfaces develop between the regions
associated with each phase, and the branchlike structures which were 
formed at the earlier stage of simulations coarsen. 
The growth of domains continues and, if left to run for a large enough 
time, the system would eventually reach the completely separated state
of two distinct layers of fluid.
Phase-separation experiments typically measure the structure factor, 
$S({\bf k},t)$, which contains information on the time evolution of 
the various length scales present in the system. It is defined as 
the Fourier transform of the density-density correlation function.
For the discrete systems we are studying, we consider equivalently
\be
S({\bf k},t) = \frac{1}{N} \left|
\sum_{\x} \left[ q(\x,t)- \bar{q}(t)\right]
e^{i{\bf k}.\x} \right|^2,
\ee
where ${\bf k}= (2\pi/L)(m{\bf i}+n{\bf j})$, $m,n=1,2,\ldots L$;
here $L$ is the linear lattice size,
$N=L^2$ is the total number of grid points in the system,
$q(\x,t)=n^{water}(\x,t)-n^{oil}(\x,t)$ 
is the total color order parameter at grid point ${\bf x}$ and 
time $t$ and $\bar{q}(t)$ is the 
spatial average of q(\x,t) at time $t$. $S({\bf k},t)$
is further smoothed by averaging over an entire 
shell in ${\bf k}$ space to obtain the circularly-averaged structure
factor 
\be
S(k,t) = \frac{\sum_{\hat{k}} S({\bf k},t) } {\sum_{\hat{k}} 1}
\ee
where the sum $\sum_{\hat{k}}$ is over a circular shell defined by 
$(n- \frac{1}{2}) \leq \frac{ |{\bf k}|L}{2\pi} < (n + \frac{1}{2})$ and
the cutoff wavevector $k_c$ has the maximum possible value
which is compatible with the periodicity in ${\bf k}$ space.
We use the first moment of the circularly-averaged
structure factor as a measure of the characteristic length scale of 
the system
\be
k(t) = \frac { \sum_{k}k S(k,t) }{\sum_{k} S(k,t)}. 
\ee
The characteristic size of the domains is then given by $R(t)=2\pi/k(t)$.  

Fig. 2 displays the temporal evolution of the circularly-averaged
structure factor obtained for a critical quench in a 
$512^2$ system.
At early times in the simulations ($t< 4000$ 
timesteps) we observed that the amplitude of the peak in the structure 
factor increases without the position of the peak changing in time.
This behavior is indicative of  initial sharpening of the
domains, as the amplitude of the peak in $S(k,t)$ 
is proportional to the domain mass within a characteristic domain
size.
As time evolves further, the peak of $S(k,t)$ shifts 
to smaller wavenumbers and its height increases, indicating 
the coarsening of the domains. At intermediate times 
we also observe the appearance of a second peak in $S(k,t)$, 
which later on becomes a shoulder of the main peak and then merges with
it. After an initial transient regime,
the dynamics of phase separation is often characterized by a single 
time scale. This length scale is usually described by a power law
behavior $R(t)= t^\alpha$.
Simple dimensional analysis of the hydrodynamical evolution equations
in $2D$ yields for the domain growth exponent $\alpha=2/3$, when the binary 
system is in the inertial hydrodynamic regime \cite{furakawa,bray}.

In Fig. 3 the time evolution of $R(t)$ is shown as obtained from our 
lattice-Boltzmann simulations with system size $512^2$. 
Finite-size effects in this quantity are known to become important when
$R(t)$ is comparable to the linear size of the lattice $L$.
We checked these effects by performing additional calculations of 
$R(t)$ for systems size $128^{2}$ and $256^2$. By comparing 
the results we deduced that finite-size errors in our $2D$ 
simulations become important when $R(t)\geq L/5$. 
The $3D$ lattice-Boltzmann simulations of Kendon {\it et al.} \cite{kendon}
seem to indicate a somewhat larger value of $R(t)$ beyond
which these errors become pronounced. This might indicate
that finite-size effects are more significant in $2D$.
Discarding both the early-time transient regime and the late-time
regime where finite-size effects are pronounced,
we found that the late-time  behavior of $R(t)$ in our simulations 
is $R(t)\sim t^{0.66\pm 0.01}$, in good agreement with the above-mentioned
scaling arguments and previous lattice-gas \cite{emerton2}
and lattice-Boltzmann \cite{alexander} simulations of phase
separation in $2D$. We note, however, that this result is only a first
qualitative study of the dynamics of phase separation within 
our model and more work is needed in order to unambiguously 
identify different scaling regimes within the parameter space of 
the  model.
\subsection{Microemulsion phases: oil droplets in water and sponge phase}
We used our model to simulate the different ternary microemulsion
phases that are possible in 2D, namely the oil-in-water droplet and 
 sponge phases. In experimental systems the two distinct
microemulsion phases
will form when the appropriate concentrations of oil, water and 
surfactant are present in the system below the critical temperature.
The oil-in-water droplet phase typically consists of finely divided spherical regions 
of oil, with stabilizing monolayers of surfactant surrounding them,
embedded within a continuous  water background. If one increases the relative 
amount of oil in the system and there is sufficient amphiphile
present, one observes the formation of mutually percolating
tubular regions of oil in water, with a monolayer of surfactant sitting at 
the interface. In both these cases, the equilibrium state does not 
correspond to complete separation of immiscible oil and water regions, 
but rather to complex structures with very different characteristic 
length scales that form as the result of the presence of amphiphile
\cite{gompsh}.

In order to reproduce the oil-in-water droplet phase, we set up a simulation 
with a uniform mixture of oil, water and surfactant with small density
fluctuations as our initial configuration. The average densities 
of oil, water and surfactant are $0.42, 0.6$ and $0.1$ respectively.
Fig. 4 displays the results. We see the rapid formation of many 
small oil-in-water droplets, whose size initially 
increases slightly, but not without limit.
This is characteristic of the experimental droplet phase.
It occurs because the free energy penalty associated with 
the existence of many oil-water interfaces in this phase,
as compared to the complete oil-water phase separation,
is compensated by the gain in free energy due to adsorption of
surfactant dipoles at these interfaces.
If  coarsening of  oil droplets was to continue 
the interface area for adsorption of surfactant dipoles would 
decrease resulting in an increase in the amount of surfactant 
in bulk water or oil. The free energy penalty for bulk 
surfactant prevents this to happen.
The concentration of the surfactant is not visible in
Fig. 4, but is high at the interface and low in oil-rich and 
water-rich regions, with the surfactant dipoles at the interface 
pointing on average from water-rich region towards the oil-reach
droplets (see also Fig. 11 and Fig. 2 in \cite{cbc}).

In order to further quantify the result we show in Fig. 5 the time
evolution of the circularly-averaged structure factor $S(k,t)$.
Once again we observe the formation of a distinct peak in the structure factor,
indicating the sharpening of the domains.
Interestingly, this happens much faster than in the case of the binary
system. The presence of the surfactant seems to accelerate  domain
formation in the early stage of phase separation, an effect which has
also been seen in the hybrid Ginzburg-Landau simulations 
of Kawakatsu {\it et al.} \cite{kawakatsu}, 
and should be detectable experimentally.
As time evolves the peak in $S(k,t)$ increases in height and shifts
further towards smaller values of the wavelength, indicating the
growth of droplets. 
From at least timestep $5000$ onwards there appears to be a negligible 
amount of further movement of the position of the peak, 
indicating that droplets have reached a maximum size and will grow 
no more. We observed small oscillations in the intensity of 
the peak in $S(k,t)$ which suggest that the characteristic domain 
mass fluctuates in time.

To investigate the ability of the model to access the sponge phase, we
set the average densities of water and oil both equal to $0.5$ while
keeping the average density of surfactant at the same 
value as before. The results are shown in Fig. 6.
Starting once again from a perturbed uniform mixture of fluids,
this time we observe the growth of an interconnected 
network of tubular regions of oil and water. 
The width of the oil and water regions 
grows in size up to about $4000$ timesteps. During this time the
surfactant particles, which were initially distributed uniformly in 
the system, concentrate around the various oil-water interfaces. 
Beyond this stage the system changes very little, indicating that 
the observed sponge phase is stable. This is also confirmed by 
our result for the circularly-averaged structure factor shown in 
Fig. 7.
\subsection{The effect of surfactant on domain growth dynamics}
To further investigate the effect of surfactant on domain growth 
dynamics of the sponge-phase we performed additional simulations 
in which we kept the average density of oil and water fixed at $0.5$ 
while gradually increasing the amount of surfactant in the system. 
The average densities of
surfactant used in these simulations are $0.05,0.1,0.15,0.20,0.30$.
We analyze the effect of varying the surfactant concentration using 
the domain size $R(t)$ as a quantitative measure. 
The domain size is calculated from the circularly-averaged structure
factor, as described in Section II.
The results are summarized in Fig. 8 where the domain sizes 
are plotted against time, and as a function of increasing surfactant 
concentration. The presence of the surfactant significantly
retards the growth of the domains and it can be clearly seen that 
for  surfactant concentrations larger than $0.05$, the domain size
reaches saturation. 
Following Boghosian, Coveney and Emerton \cite{emerton2}, we analyzed the
time-dependence of domain growth in terms of a 
stretched exponential form
\be
R(t) = R_{\infty} - a\exp(-ct^d)
\ee
where $R_{\infty}$, $a$, $c$ and $d$ are adjustable parameters,
which are determined by a least-square fit to our data.
As can be seen from Fig. 8 this form fits our results extremely well
across the full time scale
of the simulations and for all surfactant concentrations considered.
We also investigated a fit of the logarithmic form 
$ R(t)= a + b(\ln t)^c$, which describes
the phase-separation of binary alloys in the presence of 
impurities \cite{alloys}. Obviously, this form
is unable to describe the late time 
saturation of the domain size and we found that 
the  root-mean-square errors using this
form to describe the early times of domain formation are also 
an order of magnitude larger than that of the stretched exponential
form. In the case with average surfactant 
density equal to $0.05$ the domain size does not saturate, nevertheless 
Eq. (32) provides a better fit to the slow growth 
of the domain size than the logarithmic form for the 
time interval we considered ($0< t \leq 12000$).
LGA simulations \cite{emerton2,love1} indicate, however, that
the logarithmic from is a good fit for describing the dynamics of 
``metastable'' (i.e. long-lived) emulsions which do eventually phase separate.

At late times in these simulations, small but persistent oscillations in
$R(t)$ can be seen in Fig. 8, which are absent 
in the case of binary systems. These oscillations have also been observed
in LGA simulations, both  in $2D$ \cite{emerton2}, and 
in $3D$ \cite{love1}.
Their presence in our lattice-Boltzmann simulations confirms
that these oscillations are caused by the additional dynamics 
which the presence of amphiphile introduces into the system: 
they are not a consequence  of statistical fluctuations 
inherent in LGA.

Finally, we investigate the relationship between the final size of the
domains and the surfactant concentration. Laradji {\it et al.} 
\cite{laradji} used simplified MD simulations to study the dynamics of ternary 
oil, water and surfactant system and found the  final domain size
$R_f$
is inversely proportional to the average concentration of surfactant
molecules $\bar{n}_s$:
\be
R_{f} \sim \frac{1}{\bar{n}_s}.
\label{eq:final}
\ee
\noindent
In the deep quenches with no system fluctuations  
performed by Laradji {\it et al} surfactant molecules entirely reside at
the interfaces.
This was not the case in the lattice-gas simulations
of Emerton {\it et al.} \cite{emerton2} 
in which a certain amount of surfactant existed 
as monomer in bulk oil or water. After 
subtracting away from $\bar{n}_s$ a background monomer density, 
these authors also found a linear relationship 
between $R_{f}$ and $1/\bar{n}_s$.
A similar situation to lattice-gas simulations 
occurs in our simulations where a significant fraction of 
surfactant exists as monomers in bulk oil or water. However, we
found that even without correcting for the background monomer
density of surfactant Eq. (~\ref{eq:final}) gives a very good description of the 
relationship between $R_{f}$ and $\bar{n}_s$.
This is shown in Fig. 9 where $R_{f}$ is plotted versus the 
inverse of $\bar{n}_s$.
The condition that all surfactant molecules should be at the interface
does not therefore seem necessary for ~(\ref{eq:final}) to hold, as long as the
surfactant molecules are mainly concentrated at the interface.

\subsection{Ternary phase: lamellae}
Next we use our model to investigate the stability of a lamellar
structure, which is composed of alternating layers of oil-rich and 
water-rich phases separated by a layer of surfactant. A preliminary
discussion was given in \cite{cbc}. 
We look at the system with and 
without surfactant present in order for a critical comparison to be
made. A similar investigation of the lamellar structure in
$2D$ and in $3D$ has been performed previously using 
LGA \cite{emerton1,love2}. In a similar way to these simulations,
we set up the initial configuration of the system as layers of
pure oil and pure water eight sites wide such that each species
has an average density $1.0$.

Snapshots of our simulations for the binary case are shown in Fig.
10 for timesteps $t\leq 9000$.
As can be seen from this figure, the initial layered structure
becomes less sharply defined as time evolves but 
the lamellar structure remains 
intact and does not evolve to complete phase separation. 
By letting the simulations run for much longer times ($30000$
timesteps), we checked that
the lamellar structure is indeed the final equilibrium state of the
system and we are not observing a metastable state with long
equilibration time. 
We also examined the stability of the structure against small 
fluctuations in the density and found the lamellar structure to remain stable.
These results are in sharp contrast with the previous LGA simulations 
in which, starting from a layered structure, 
complete phase separation was observed \cite{emerton1}.
As pointed out in \cite{cbc}, in an infinite two-dimensional system 
with finite-temperature fluctuations, one expects lamellar structures
to be unstable, due to Peierls instabilities \cite{peierls}. 
The Peierls theorem, however, does not make any statement about the 
stability of such structures in finite $2D$ systems.
The stability of the lamellar structure seen in our lattice-Boltzmann
simulations and its instability in LGA simulations thus provides numerical 
evidence that the Peierls mechanism is also capable of destabilizing
periodic layered structures in finite systems.

Next we examine the effect of surfactant on the layered structure by
setting up a simulation where there is, in addition to water and 
oil layers, a single layer of surfactant at each of
the oil-water interfaces. The
average density of surfactant in each monolayer was $1.0$ and 
the initial condition for the surfactant dipole vectors at each site 
was ${\bf d}(\x,0)=0$.
We found that the final state of the system is, once again, lamellar.
The presence of surfactant, however, 
causes the water-oil interfaces to
remain sharper than in the previous simulations. This effect is shown 
in Fig. 11 (top panel) where the initial and final state of the 
color order parameter (averaged over the $y$ direction, the direction 
perpendicular to layers) are plotted along the $x$-axis for both sets of
simulations. 

Due to the symmetry of the system $d_y$, the component of the dipole
vector in the $y-$direction, does not evolve in time from 
the initial condition $d_y=0$. Under, the influence of the 
field set up by the color gradient, however, $d_{x}$ does evolve in time. 
A plot of the final state of $d_x$, averaged in the $y$ direction, is 
shown in figure 11 (lower panel). It can be seen that the surfactant directors
have their largest values around the interfaces, i.e. at the points where
the color order parameter changes sign. 
It is interesting, albeit expected, that the surfactant dipoles alter
alignment, always pointing from water-rich to oil-rich layers. 
The reason for this behavior is 
that, neglecting  surfactant-surfactant interactions,
the direction of the equilibrium dipole vectors in our model 
is determined by the gradient of the color order parameter, as can be seen by 
expanding Eq. (24) in a Taylor series in the ratio of $|{\bf c}_i$ to
the color gradient scale length.
For the lamellar structure, the color order parameter changes only in the
$x-$directions with its slope changing sign alternately,  
in passing from a water-rich layer to an oil-rich layer, as can be seen in 
Fig. 11 the surfactant dipole vectors flip direction every time this
happens.
\subsection{Self-assembly in mixtures of water and surfactant}
Repeating the simulations performed for the ternary mixtures 
but setting the average density of oil (or water) to zero gives 
results for binary water and surfactant system.
We kept the  average density of water fixed at $0.5$ and 
performed two simulations, one with a high surfactant density
$\bar{n}_s=0.25$ and one with a low surfactant density
$\bar{n}_s=0.071$. Snapshots of the simulation for the case 
of high surfactant concentration are shown in Fig. 12.
It can be seen that starting from 
a uniform mixture of water and surfactant, the system organizes 
itself in small domains each of which has a clear lamellar structure
consisting of alternating water-rich and surfactant-rich 
layers.
These domains grow in time but not without limit. 
They are highly dynamic objects which
continuously form and disintegrate but whose average size does not
grow in time once they are formed. 
In Fig. 13 we display the variation of water and surfactant densities
within one of the domains, as an example.
The densities are averaged over the $y-$direction
within the domain and are displayed along the $x-$axis (the direction
of density modulations within this domain). Also shown are  $d_x$ and $d_y$
components of  surfactant directors. It can be  seen
that the domain is built up of a stack of surfactant-rich bilayers 
separated by water-rich layers each $\sim 2$ lattice units wide. 
The surfactant directors are ordered anti-ferromagneticaly within
each domain, such that only the hydrophilic heads of surfactant 
molecules are exposed to water-rich regions.
In the case of the system with low surfactant density  visualization 
of the data indicates the existence of weak density modulations 
in the system but without any clear domain formation.

To further quantify the dynamics 
of self-assembly in the binary water-surfactant system 
we make use of the circularly-averaged
surfactant structure factor. In figure 13 we show $S(k,t)$ 
at timesteps $0$, $1000$ and $2500$ for both  systems.
It can be seen that in both cases $S(k,t)$ has a peak around 
$k=1.6$. This peak corresponds to the repeat period of water and 
surfactant density modulations and its position becomes 
stable already for $t< 100$ timesteps. In the case of the system with high 
surfactant concentration we see the emergence of a second peak in 
$S(k,t)$ at much smaller wavectors, indicating that there is 
another characteristic length-scale in this system, namely the 
average size of lamellar domains. 

In order to identify 
the driving force behind self-organization of the system we 
performed two additional calculations, using the same initial 
conditions as before. One simulation was performed 
with dipolar interactions among surfactant molecules 
switched off ($g_{ss}=0$) while in the other simulation we kept 
$g_{ss}=0.01$ but switched off the coupling between surfactant and 
water molecules, by setting $g_{\sigma s}$ to $0$. In Fig. 14 the spherically 
averaged structure factor is shown at timestep $2500$ as obtained 
from these simulations. For comparison $S(k,t)$ of the full 
simulations is also shown at the same timestep.
It can be seen that switching off the interaction between surfactant 
dipoles result in a structure factor which has only a single 
peak near $k=1.6$ while switching off water-surfactant interaction
results in a structure factor with only a peak near $k=0$.
This provide clear evidence that water-surfactant interactions
are responsible for the formation of alternating water-rich and 
surfactant-rich layers while formation of domains is
a consequence of dipolar interactions between surfactant particles.

\section{Conclusions}
Building on our recent work \cite{cbc} we developed in this paper
a lattice-Boltzmann model for ternary interacting amphiphilic fluids.
The main features of the model are that interactions among fluid components 
are realized by introducing self-consistently 
generated mean-field forces between different species and that 
the orientational degrees of freedom of amphiphilic species are 
explicitly modeled.
The  mean-field force is incorporated into the scheme in a way which 
is consistent with the kinetic theory of interacting fluids mixtures
and we provided a physical interpretation for the action of this 
force in terms of introducing 
off-diagonal matrix elements in  the BGK collision operator.

Using a single set of force coupling constants, we have shown that our
model exhibits
the correct $2D$ phenomenology for both binary and ternary phase systems.
Various experimentally observed self-assembling structures form in a
consistent way as a result of altering the relative amounts of oil,
water and amphiphile in the system. The presence of enough surfactant
clearly arrests the growth of the domains and we showed that  when 
this happens the final domain size is inversely proportional to the
amount of surfactant present in the system, in agreement with
previous molecular dynamics simulations. Our study of the stability 
of the lamellar structure in $2D$ confirmed 
a striking difference between lattice-gas and lattice-Boltzmann
simulations which result from the
absence of fluctuations in the lattice-Boltzmann scheme \cite{cbc}.
Self-assembly into local lamellar structure, as found in our simulations, 
has not been reported previously for binary water-surfactant systems
but has been observed in microemulsion experiments performed on 
ternary water, oil and surfactant systems\cite{vonk}.
Our additional calculations indicate that by increasing 
the force coupling constants beyond a certain value 
the global lamellar phase, which has been reported previously 
\cite{gompsh,matsen}, can also be reached in our model.
The ability of our model to simulate
self-assembly of surfactant aggregates in a binary water-surfactant 
system clearly distinguishes our model from other 
existing lattice-Boltzmann schemes \cite{julia1,gompers1,gompers2}
which do not incorporate the vectorial nature of surfactant
molecules and are therefore unable to describe the formation 
of such aggregates.

Natural refinements of our model would be the inclusion of
fluctuations, e.g, via a fluctuating force term compatible 
with the fluctuation-dissipation theorem \cite{ladd} 
which enables the system to move out of its metastable states. 
Also, some of the instabilities which we encountered in the 
present model might be mitigated by using more realistic forms 
for the interaction between different molecules and by adding 
the Enskog corrections for the collisions in dense systems \cite{enskog}
to the BGK collision operator.  
Our recently developed  parallel
\cite{jon} version of the lattice-Boltzmann model 
should allow us to extend the present
study to $3D$ for which much more experimental data is available.
This will help us to select model parameters so that  
the model will provide a more realistic description of experimental 
observations; we also hope to then study various important phenomena
such as flow of complex fluids in porous media.

\section*{Acknowledgments}
M.N. and P.V.C. thank Peter Love, N\'{e}lido Gonz\'{a}lez-Segredo,
J. Chin and  Rammile Ettelaie
for useful discussions. This work was partially supported by EPSRC
under Grant No GR/M56234. P.V.C, B.M.B and H.C. are indebted to NATO 
for a travel grant (Grant No. CRG950356) which facilitated 
their collaboration. 

\newpage
 \begin{figure}\noindent
\caption{Snapshots of 
phase separation of a binary oil-water mixture
during a critical quench.
From left to right and top to bottom
timesteps $0,800,1600,
2400,3200,4800,6400,8000,12000,16000,20000,24000,30000,,45000,60000,90000$
are shown. The system size is $256^2$.} 
\end{figure}
\begin{figure}\noindent
\caption{Temporal evolution of $S(k,t)$ for a critical quench 
in a binary oil-water system. 
Timesteps shown are, from bottom to top,
$t=3200,4000,6400,9600,12000,14400,20000,24000$. The system size is $512^2$.}
\end{figure}

\begin{figure}\noindent
\caption{Logarithm of the average domain size $R(t)$ (lattice units) 
as a function of the logarithm
of the time (timesteps) with data taken from lattice-Boltzmann simulations of a
critical quench in binary immiscible phase separation. 
The straight line corresponds to a growth exponent 
$\alpha=0.66$ and is provided as a guide to the eye only. 
The system size is $512^2$.}
\end{figure}

\begin{figure}\noindent
\caption{Snapshots of time evolution of oil-in-water microemulsion phase. From 
left to right and top to bottom
timesteps $0,200,400,600,800,1200,1600,2000,3000,4000,5000,6000$ 
are shown. The system size is $256^2$.}
\end{figure}
\begin{figure}
\caption{Temporal evolution of the circularly-averaged structure factor
$S(k,t)$ for the microemulsion droplet 
phase shown in Figure 4. Timesteps shown, from bottom to top,
are $t=200,1000,2000,3000,5000,6000$.}
\end{figure}

\begin{figure}\noindent
\caption{Snapshots of time evolution of sponge microemulsion phase. 
From 
left to right and top to bottom
timesteps $0,200,400,600,800,1200,1600,2000,3000,4000,5000,6000$ 
are shown. The system size is $256^2$.}
\end{figure}

\begin{figure}
\caption{Temporal evolution of the circularly-averaged structure factor
$S(k,t)$ for the sponge microemulsion phase shown in Figure 6.
Timesteps shown, from bottom to top,
are $t=200,1000,2000,3000,4000,6000$.}
\end{figure}

\begin{figure}\noindent
\caption{Time evolution of average domain size $R(t)$ (lattice units)
in a ternary system with equal concentrations of water and oil ($0.5$).
Curves from top to bottom correspond to systems with average surfactant
concentration $0.05,0.1,0.15,0.2,0.3$. The full lines are the 
stretched exponential fits to our data.}
\end{figure}

\begin{figure}\noindent
\caption{
Final domain sizes $R_{f}$ (lattice units) as a function of the inverse of the
surfactant concentration, $1/\bar{n}_s$. The solid line is a linear
fit to the first $4$ points of our data.}
\end{figure}

\begin{figure}[h]\noindent
\caption{Snapshots of the evolution of the lamellar structure in the
absence of surfactant. Timesteps shown are clockwise from top to bottom 
$t=0,300,600,3000,6000,9000$. The system size is $256^2$.} 
\end{figure}

\begin{figure}\noindent
\caption{Upper panel shows the final state color order parameter, 
averaged over the $y-$direction (vertical in Fig. 10) of lamellar structure, 
both with and without surfactant present. The timestep shown is $9000$.
Note how the presence of surfactant sharpens the interfaces between
water and oil.  
Lower panel shows the final state distribution of the surfactant directors, 
averaged over the $y-$ direction at timestep $9000$.
The system size is $256^2$.}
\end{figure}

\begin{figure}\noindent
\caption{Snapshots of self-assembly of a uniform mixture 
of water  and surfactant into lamellar domains. From left 
to right and top to bottom timesteps $0$, $300$, $3000$, $9000$. The
surfactant-water density difference is shown in grey scaling 
with black corresponding to high surfactant density and white 
corresponding to high water density.
The average concentrations
of water and surfactant are $0.5$ and $0.25$, respectively.
The system size is $256^2$. }
\end{figure}

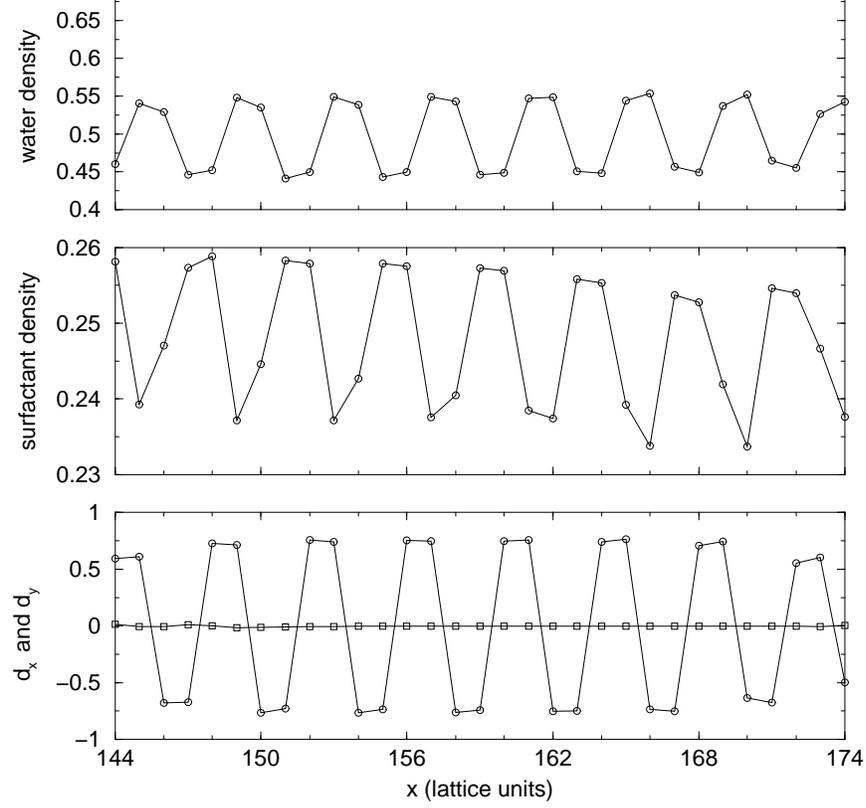
\begin{figure}\noindent
\caption{Upper panel shows variation  of the water density within 
an ``anti-ferromagnetic'' domain averaged over the $y-$direction (which 
is vertical in figure 12) within the domain. 
Middle panel shows variation of surfactant density within the same
domain.
Lower panel shows the final state distribution of the $d_x$ (circles)
and $d_y$ (squares) of surfactant directors, 
averaged over the $y-$ direction. The timestep shown in all panels is 
$3000$.}
\end{figure}

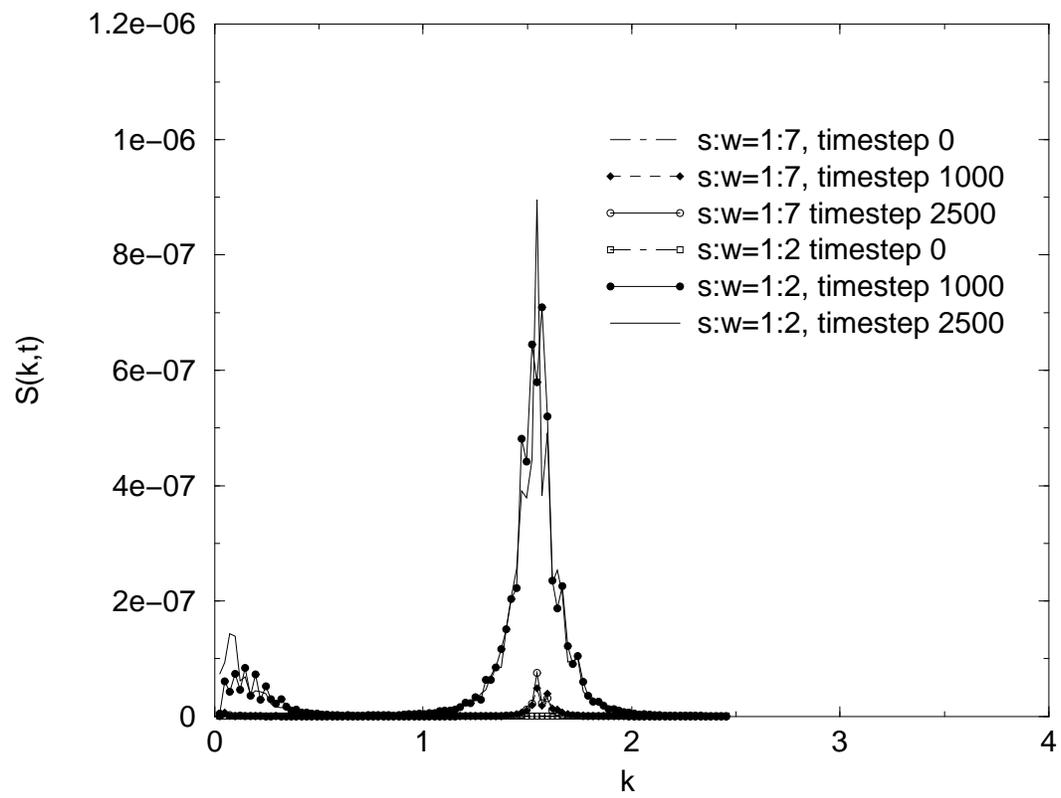
\begin{figure}\noindent
\caption{Circularly-averaged surfactant density
    structure factor $S(k,t)$ for binary water and surfactant
    mixtures shown at timesteps $0$ and $2500$ for a 
    $256^2$ binary water-surfactant system.
    The average density of water is kept at $0.5$ while the
    surfactant average density is increased from $0.071$ 
    s:w=1:7), to  $0.25$ (s:w=1:2). For both systems $S(k,t)$ 
    has a  peak near $k=1.6$ whose position corresponds to the 
    repeat period of the lamella.
    The peak near $k=0$ is present only for the system with 
    high surfactant concentration and it signals the formation of 
    lamellar domains whose average size correspond to the position 
    of this peak.}
\end{figure}

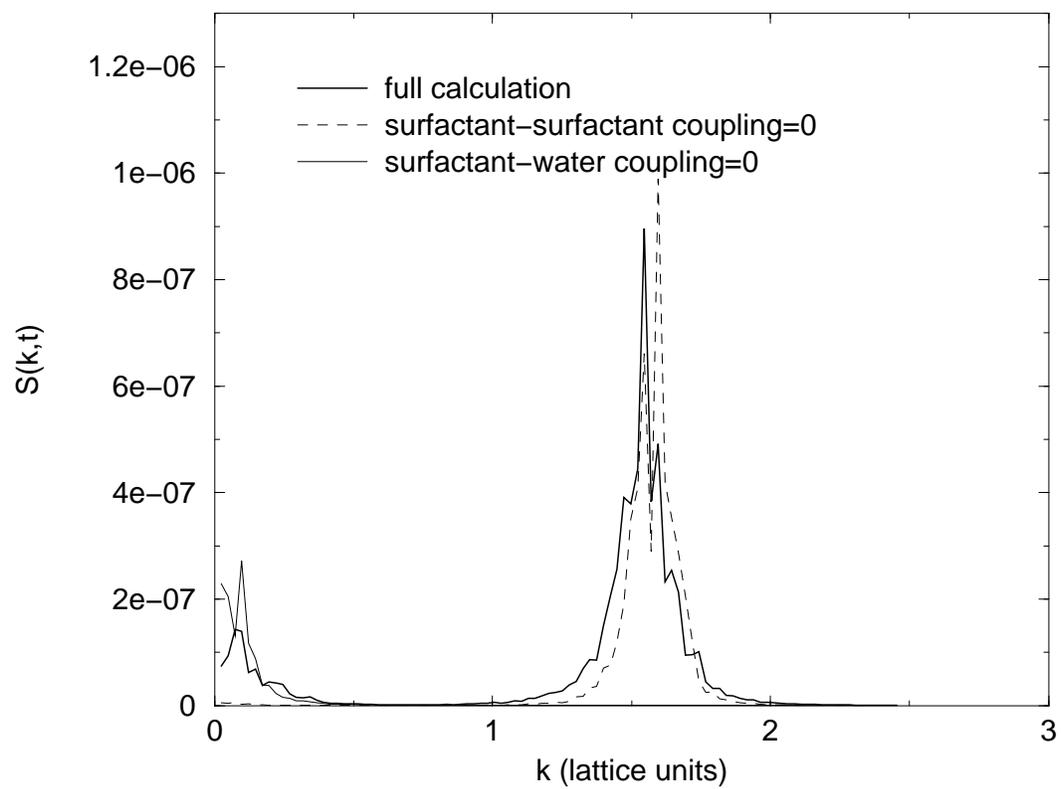
\begin{figure}\noindent
\caption{Circularly-averaged surfactant density
    structure factor $S(k,t)$ for a binary water and surfactant
    mixture at timestep $2500$ calculated 
    with both water-surfactant and surfactant-surfactant 
    coupling swtiched on (thick solid line), 
    only surfactant-water coupling switched on (dashed line) and
    only surfactant-surfactant coupling switched on (solid line).
    The system size is $256^2$.
    The average concentrations  of water and surfactant are 
     $0.5$ and $0.25$ respectively.}
\end{figure}

\end{document}